\documentclass{article}

\title{The spinor equation for the electromagnetic field}

\author{Baoxia Su}
\begin{document}
\maketitle
\centerline{Department of Applied Physics, Tianjin University, 
P. R. China}
\bigskip
{\bf\centerline{ABSTRACT}}
     We develop a spinor equation of the electromagnetic field, 
which is 
equivalent to the Maxwell equation and has a similar form as the 
Dirac 
equation. The spinor is the very conjugate momentum of the vector 
potential
in the Lagrangian mechanics. In this framework the electromagnetic field described
by the spinor exhibits the SU(2) internal symmetry. The quantization and the 
problem of longinal and scalar photons are disscussed.

\begin{flushleft}
PACS number(s): 41.20.Bt, 03.50.De

\bigskip
{\bf Introduction}
\end{flushleft}

    Conventionally, to describe the electromagnetic field one has 
introduced the electromagnetic tensor as well as the vector potential. 
However, the tensor has some redundant components and the form of the 
field equation is not as simple and compact as the Dirac equation, 
which can not only give us inconvenience in applications but also hinder
us from revealing the internal symmetry of the system.

   In this paper we introduce a spinor in replace of the previous tensor 
to describe the electromagnetic field, by the help of which the Maxwell 
equation is equivalent to the spinor equation which has the similar form 
as the Dirac equation. The spinor happen to be the conjugate momentum of
the vector potential. We also exploit the internal symmetry of the field.

In section I we give the classical form of the spinor equation of the field.
In section II we study the Fourier property of the free field. In section III we showed the Lorentz invariance of the equation. In section IV we give the Lagrangian of the field and  study the symmetry. In Section V we study the quatization of the field in the Lagrangian mechanics.        

\section{The new wave equation for the electromagnetic field}

 As is know that the Maxwell equation can be expressed in the complex
 form as
$$\matrix{\nabla\times\stackrel{\rightarrow}{\phi}-\partial_4\stackrel{
\rightarrow}{\phi}=\sqrt{\frac{\mu_0}{2}}\stackrel{\rightarrow}{j}_e\cr
  \nabla\cdot\stackrel{\rightarrow}{\phi}=\sqrt{\frac{\mu_0}{2}}j_{e4}}
\eqno{(1.1)}$$
where
$$\stackrel{\rightarrow}{\phi}=\frac{1}{\sqrt{2\mu_0}}(\stackrel
{\rightarrow}{B}+i\stackrel{\rightarrow}{E})$$
is a three-component vector in the coordinate space, and $j_e$ is the 
four-component electric current vector.

If we consider $\phi$ a three-component column and add a fourth 
component 
to it, we get a four-component column
$$\phi\equiv\pmatrix{\stackrel{\rightarrow}{\phi}\cr \xi}\eqno{(1.2)}$$
Using Eq.(1.1) we can construct a wave equation of $\phi$ defined in Eq.(1.2) 
as
$$\Lambda_\mu \partial_\mu \phi=\sqrt{\frac{\mu_0}{2}}j_e\eqno(1.3)$$
where

$$\matrix{\Lambda_1=\pmatrix{0&0&0&-1\cr 0&0&-1&0\cr 0&1&0&0\cr 
1&0&0&0},&
\Lambda_2=\pmatrix{0&0&1&0\cr 0&0&0&-1\cr -1&0&0&0\cr 0&1&0&0},\cr
\Lambda_3=\pmatrix{0&-1&0&0\cr 1&0&0&0\cr 0&0&0&-1\cr 0&0&1&0},&
\Lambda_4=-I}$$
are unitary matrices.

   Considering that $j_e$ is a conserved current of the Dirac 
field, we get that $\xi$ satisfys the Klein-Gordon eqation 
$$\partial_\mu\partial_\mu\xi=0.\eqno{(1.4)}$$
As $j_e$ is a Lorentz vector $\xi$ can be proved to be a 
Lorentz scalar. Therefore $\xi$ can have a zero solution 
$$ \xi(\stackrel{\rightarrow}{x},t)=0\mbox{ (for any $\stackrel{
\rightarrow}{x}$ and t)}.\eqno{(1.5)}$$ 
Apperantly, Eq.(1.3) is compeletly equivalent to the Maxwell 
equation
(1.1) under the condition Eq.(1.5). We adopt this condition 
in this and
 next section. Eq.(1.5) will be called the the transversality 
condition, 
which is the requirement of Maxwell equation to Eq.(1.3).

   One can easily check that $\phi$ can be described in terms of the 
electromagnetic potential A as
$$\phi=-\frac{1}{\sqrt{2\mu_0}}\Lambda_\mu^{-1}\partial_\mu A\eqno{(1.6)}$$
From Eq.(1.6) we know that Eq.(1.5) produces the Lorentz condition
$$\partial_\mu A_\mu=0.\eqno{(1.7)}$$
The usual four-component electromagnetic energy flux density can be expressed as
$$j_\mu =-i\phi^+ \Lambda_\mu\phi$$

Therefore the Maxwell equation is now descriped in Eq.(1.3) plus the condition 
Eq.(1.5) ,which is equivalent to Lorentz condition.

\section{The Fourier decomposition of the free field}

 In this section we discuss the properties of the free field on the condition Eq.(1.5). we will give the plane wave solution and show the Lorentz invariance of the field.  

For free field Eq.(1.3)reduces to
$$\partial_\mu \Lambda_\mu \phi=0\eqno{(2.1)}$$
which has the similar form as Dirac Equation.

The plane wave solution to Eq.(2.1) can be expressed as
$$\phi_{\stackrel{\rightarrow}{k},\alpha}(\stackrel{\rightarrow}{x},t)=
C_{\stackrel{\rightarrow}{k},\alpha}u_{\stackrel{\rightarrow}{k},
\alpha}e^{i(\stackrel{\rightarrow}{k}\cdot \stackrel{\rightarrow}{x}-\omega t)}
\eqno{(2.3)}$$
where $C_{\stackrel{\rightarrow}{k},\alpha}$ is a constant,
 $u_{\stackrel{
\rightarrow}{k},\alpha}$ is a unit four-component column, satisfying
$$u_{\stackrel{\rightarrow}{k},\alpha}^+ u_{\stackrel{\rightarrow}{k},
\alpha^\prime}=\delta_{\alpha,\alpha^\prime}$$
and $\alpha$ is the energy sign parameter, satisfying
$$\omega=\alpha\mid \stackrel{\rightarrow}{k}\mid  \eqno{(2.4)}$$
where $\alpha=\pm 1$. Any electromagnetic field can be expanded on the basis of the plane wave solution as in Eq.(2.3).

By virtue of Eq.(1.5) one obtains
$$\matrix{u_{\stackrel{\rightarrow}{k},\alpha,4}=0\cr
\stackrel{\rightarrow}{k}\cdot \stackrel{\rightarrow}{u}_{\stackrel{
\rightarrow}{k},\alpha}=0}\eqno{(2.5)}$$
which states that the fourth component of $u_{\stackrel{\rightarrow}{k},
\alpha}$ is zero and the first three components are transverse. 
Therefore 
 the condition (1.5) as well as the Lorentz condition(1.7) is the 
transversality condition. 
We should note that without the transversality condition Eq.(2.1) has 
other solutions that do not satisfy Eq.(2.5) and can be viewed as the 
longinal and scalar field.

\section{The spinor equation and the helicity of the combined field }

In this section we drop the constraint condition Eq.(1.5) attached to
Eq.(1.3) and see what we get. We can suppose the Eq.(1.3) describe the combined field: electromagnetic field plus a scalar field.
After careful check we will find the $\phi$ defined in Eq.(1.2) is a spinor and
$\xi$ is a scalar.

   Considering that $j_e$ in Eq.(1.3) is a Lorentz vector, we can prove that 
when the system undergoes the normal Lorentz transformations, $\phi$ 
transforms as
$$\phi^\prime=T\phi$$
where
$$T=-\Lambda_\mu^{-1}a_{\mu4}a\eqno{(3.1)}$$
and the form of Eq.(1.3) keeps invariant in the new frame. Therefore 
$\phi$ is a spinor and Eq.(2.1) is the spinor equation for the free 
combined field, which is comparable to Dirac equation. 
From Eq.(3.1) one can get that $\xi$ is a scalar, which provides the 
justification for the condition (1.5).
   
From Eq.(3.1) the spin matrices are found to be
$$s_1=\pmatrix{0&0&0&0\cr 0&0&-i&0\cr 0&i&0&0\cr 0&0&0&0},
  s_2=\pmatrix{0&0&i&0\cr 0&0&0&0\cr -i&0&0&0\cr 0&0&0&0},
  s_3=\pmatrix{0&-i&0&0\cr i&0&0&0\cr 0&0&0&0\cr 0&0&0&0}\eqno{(3.2)}$$
,which indicates that the field has spin 1. 

The field helicity can be defined as
$$h\equiv\stackrel{\rightarrow}{s}\cdot \stackrel{\rightarrow}{p}/\mid 
\stackrel{\rightarrow}{p}\mid$$
which is conserved in time. We find that the plane wave as in Eq.(2.3) with 
the condition (2.5) has the  $-\alpha$ helicity
$$h\phi_{\stackrel{\rightarrow}{k},\alpha}=-\alpha\phi_{\stackrel{
\rightarrow}{k},\alpha}\eqno{(3.3)}$$
We should note that, without the constraint condition Eq.(1.5), Eq.(2.1) has a zero 
helicity solution $\phi_{\stackrel{\rightarrow}{k},\alpha}^l$ satisfying
$$h\phi_{\stackrel{\rightarrow}{k},\alpha}^l=0.$$

We must note that $\phi_{
\stackrel{\rightarrow}{k},\alpha}^l$ is beyond the extent of the 
actual electromagnetic field. In fact one can check that $\phi_{
\stackrel{\rightarrow}{k},\alpha}^l$ represents the longinal or scalar field 
while $\phi_{\stackrel{\rightarrow}{k},\alpha}$ represents the transverse 
field. It is the Maxwell equation that rules out the possibility of longinal 
and scalar fields.

In this section we show that Eq.(1.3) is invariant under 
Lorentz transformation and $\phi$ is a spinor.

\section{The Lagrangian and the internal symmetry of the field}
  
In fact Eq.(1.5) is the classic express of the transversality condition. In this section in order to prepare for the quantazition
 we drop the constraint condition Eq.(1.5). We study some properties 
of the combined field without the condition(1.5) in Lagrangian 
mechanics in this section. 

  We seek a genaralized potential A which is free of lorentz condition and satisfys Eq.(1.6). The fourth component
in Eq.(1.6) reads 
$$ \xi=\frac{1}{\sqrt{2\mu_0}}\partial_\mu A_\mu. \eqno{(4.1)}$$
From Eq.(1.4) we know that the field $\xi$ is a massless Klein-Gordon field. 
The Lorentz condition as in Eq.(1.7) no longer holds.

The Lagrangian density of the combined field can be defined by
$$\textit{L}=-\phi^2\eqno{(4.2)}$$
We suppose A is the canonical coordinate of the field and we get the conjugate momenta 
$$\pi_\mu=i\sqrt{\frac{2}{\mu_0}}\phi_\mu.$$
Therefore $\phi$ is the conjugate momenta of the canonical coordinate A
 except for a factor. The Hamiltonian density is found to be
$$\textit{H}=\frac{1}{\sqrt{2\mu_0}}\phi\Lambda_\mu \partial_\mu A.\eqno{(4.3)} $$
 By studying Eq.(4.2) one can find the system in free state has an internal 
symmetry. There exists an isospin operator 
$$\stackrel{\rightarrow}{s}^{iso}=\frac{i}{2}\stackrel{\rightarrow}
{\Delta}$$
where
$$\matrix{\Delta_1=\pmatrix{0&0&0&1\cr 0&0&-1&0\cr 0&1&0&0\cr
 -1&0&0&0},&
\Delta_2=\pmatrix{0&0&1&0\cr 0&0&0&1\cr -1&0&0&0\cr 0&-1&0&0},\cr
\Delta_3=\pmatrix{0&-1&0&0\cr 1&0&0&0\cr 0&0&0&1\cr 0&0&-1&0}}$$
are unitary matrices. The corresping currents read 
$${j_\mu}^i=\pi{\Lambda_\mu}^{-1}\Delta_iA$$
which is conserved for the free field.
However the symmetry will break on the interaction with 
other fields such as the Dirac field. 

\section{The quantization of the field}

In this section we study the quatization of the combined field in the framework of last section.
In order to deal with the quantization of the field we 
introduce a set of normal orthogonal basis in the four-dimension space
$$\matrix{\epsilon_{\stackrel{\rightarrow}{k}1}=\pmatrix{\stackrel{
\rightarrow}{e}_1\cr0},&\epsilon_{\stackrel{\rightarrow}{k}2}=\pmatrix{
\stackrel{\rightarrow}{e}_2\cr 0},&\epsilon_{\stackrel{\rightarrow}
{k}3}=
\pmatrix{\stackrel{\rightarrow}{k^0}\cr0},&\epsilon_{\stackrel{
\rightarrow}{k}4}=\pmatrix{\stackrel{\rightarrow}{0}\cr 1}}$$
where $\stackrel{\rightarrow}{e}_1,\stackrel{\rightarrow}{e}_2,
\stackrel{
\rightarrow}{k^0}$ are normal orthogonal basis in the three-dimension space.
Considering that $\stackrel{\rightarrow}{A},iA_4$ are Hermitian 
operator,
 we need to define an accompanying set of basis as
$$\matrix{\stackrel{-}{\epsilon}_{\stackrel{\rightarrow}{k}i}=
\epsilon_{\stackrel{\rightarrow}{k}i}\cr
\stackrel{-}{\epsilon}_{\stackrel{\rightarrow}{k}4}=
-\epsilon_{\stackrel{\rightarrow}{k}4} }$$
for i=1,2,3. The canonical coordinate operator A can be expressed as 
$$A(\stackrel{\rightarrow}{x},t)=\sqrt{\frac{\mu_0}{2V}}\sum_{\stackrel{
\rightarrow}{k}}\sum_{\lambda=1}^4\frac{1}{\sqrt{\mid\stackrel{
\rightarrow}{k}\mid}}[\epsilon_{\stackrel{\rightarrow}{k}\lambda}
{a(t)}_{\stackrel{
\rightarrow}{k}\lambda}e^{i\stackrel{\rightarrow}{k}\cdot\stackrel{
\rightarrow}{x}}+\stackrel{-}{\epsilon}_{\stackrel{\rightarrow}{k}
\lambda}{a(t)}^+_{\stackrel{\rightarrow}{k}\lambda}e^{-i\stackrel{\rightarrow}{k}
\cdot\stackrel{\rightarrow}{x}}]\eqno{(5.1)}$$
where $a(t)_{\stackrel{\rightarrow}{k}\lambda}$ is the annihilation operator,satisfying
$$[a(t)_{\stackrel{\rightarrow}{k}\lambda},a(t)^+_{\stackrel
{\rightarrow}{k^\prime}\lambda^\prime}]=\delta_{\stackrel{\rightarrow}
{k^\prime},\stackrel{\rightarrow}{k}}\delta_{\lambda^\prime,\lambda}
.\eqno{(5.2)}$$
From Eq.(1.6) and Eq.(5.1) we get the momenta operator 
$$\phi(\stackrel{\rightarrow}{x},t)=\frac{i}{\sqrt{2V}}\sum_{\stackrel{
\rightarrow}{k}\lambda}\sqrt{\mid\stackrel{\rightarrow}{k}
\mid}[v_{\stackrel{\rightarrow}{k}\lambda}{a(t)}_{\stackrel{
\rightarrow}{k}\lambda}e^{i\stackrel{\rightarrow}{k}\cdot\stackrel{
\rightarrow}{x}}-\stackrel{-}{v}_{\stackrel{\rightarrow}{k}
\lambda}{a(t)}^+_{\stackrel{\rightarrow}{k}\lambda}e^{-i\stackrel{\rightarrow}{k}
\cdot\stackrel{\rightarrow}{x}}]\eqno{(5.3)}$$
where
$$\matrix{v_{\stackrel{\rightarrow}{k}\lambda}=-\frac{1}{\sqrt{2}\mid
\stackrel{\rightarrow}{k}\mid}k_\mu\Lambda^{-1}_\mu\epsilon_{\stackrel{\rightarrow}{k}\lambda}\cr
\stackrel{-}{v}_{\stackrel{\rightarrow}{k}\lambda}=-\frac{1}{\sqrt{2}
\mid
\stackrel{\rightarrow}{k}\mid}k_\mu\Lambda^{-1}_\mu\stackrel{-}{\epsilon}_{\stackrel{\rightarrow}{k}\lambda}}$$

From Eq.(5.1),Eq.(5.2) and Eq.(5.3) we can get that the canonical coordinate A 
and 
it's momenta $\pi$ satisfy the standard commutation relation
$$[A_\mu(\stackrel{\rightarrow}{x},t),\pi_\nu(\stackrel{\rightarrow}
{x^\prime},t)]=i\delta_{\mu\nu}\delta^3(\stackrel{\rightarrow}{x}-\stackrel{\rightarrow}{x^\prime})$$

From Eq.(4.3),(5.1) and (5.3) we get the Hamiltonian of the combined
 field

$$\textit{H}=\sum_{\stackrel{\rightarrow}{k}}\mid\stackrel
{\rightarrow}{k}
\mid[\sum_{\lambda=1}^2(N_{\stackrel{\rightarrow}{k}\lambda}
+\frac{1}{2})
+N_{\stackrel{\rightarrow}{k}3}-N_{\stackrel{\rightarrow}{k}4}]
\eqno{(5.4)}$$
where $N_{\stackrel{\rightarrow}{k}\lambda}$ is number operator.
The momentum operator of the field can be calculated as well 
$$\stackrel{\rightarrow}{P}=\sum_{\stackrel{\rightarrow}{k}}
\stackrel{\rightarrow}{k}[\sum_{\lambda=1}^2(N_{\stackrel{\rightarrow}
{k}\lambda}+\frac{1}{2})+N_{\stackrel{\rightarrow}{k}3}-N_{\stackrel{\rightarrow}{k}4}]\eqno{(5.5)}$$ 
 Eq.(5.4) and Eq.(5.5) is consistent with the conventional result.
The problem of negative energy as in Eq.(5.4) will be solved as follow.

With the help of Eq.(4.1) and (5.1) we get the fourth component 
of $\phi$ in Eq.(5.3)
$$\xi=\frac{i}{\sqrt{2V}}\sum_{\stackrel{
\rightarrow}{k}}\sqrt{\mid\stackrel{\rightarrow}{k}
\mid}[b_{\stackrel{\rightarrow}{k}}e^{i\stackrel{\rightarrow}{k}\cdot
\stackrel{\rightarrow}{x}}-h.c.]$$
,where we define a new annihilation operator as
$$b_{\stackrel{\rightarrow}{k}}=\frac{1}{\sqrt{2}}(a_{\stackrel
{\rightarrow}{k}3}+ia_{\stackrel{\rightarrow}{k}4})$$
Now we turn back to the transversality codition (1.5). The $\phi$ in Eq.(1.5) 
should be interpreted as the average of the operator in a state.  The condition
 (1.5) turns to be
$$<a\mid \xi \mid a>=0, \eqno{(5.6)}$$
where $\mid a>$ is an arbitrary state. 
 Eq.(5.6) implys that the operator b should annihilate any state 
i.e.
$$b_{\stackrel{\rightarrow}{k}}\mid a>=0\eqno{(5.7)}$$
With the constraint condition 
Eq.(5.7), $N_{\stackrel{\rightarrow}{k}3}-N_{\stackrel{
\rightarrow}{k}4}$ in Eq.(5.4) and (5.5) have no actual contribution.

In summary, the electromagnetic field described by Eq.(1.3) is 
well quantized with the condition Eq.(5.6) or Eq.(5.7).

\begin{flushleft}CONCLUSION\end{flushleft}
We rewrite the Maxwell equation into an equivalent spinor equation 
which is 
Lorentz invariant and has a Dirac-like form. The condition that 
the fourth
component of the spinor is zero or the transversality condition is necessray
and feasible, which is equivalent to the Lorentz condition. On quantization
 in the lagrangian mechanics, the fourth component is a nonvanishing operator
and the transversality condition reduces to the vanishing of the 
average of
the operator in all states. 
  The appearance of the longinal and scalar fields are 
the requiremnt of the symmetry of the field and the null observable contribution 
of them is the result of the transversality condition. 
  The spinor also play a 
definite role in Lagrangian mechanics: it is the conjugate momentum 
of the vector
potential, so the spinor and the vector are two corresponding and necessary parts 
of the field. From the Lagrangian of the field we find there exists the SU(2) 
symmetry of the field and we calculate the conserved currents. 
The field is well depicted in a symmetric and simple form by the 
employ of the 
spinor in replace of the tensor, which can not only simplify 
the calculation but also reveal the internal symmetry, and which 
will find more 
application.


\bigskip
\begin{flushleft}REFERENCES\end{flushleft}
[1]S. N. Gupta, Quantum Electrodynamics (Gordan and Breach Science, 
New York,1977 ).

\noindent [2]V. Bargmann and E. P. Wigner (1948) Proc. Roy. Acad. Sci. (USA) 34.

\end{document}